\RequirePackage[2020-02-02]{latexrelease}
\documentclass[aps,prl,showpacs,twocolumn]{revtex4}

\usepackage{amssymb}
\usepackage{graphicx}
\usepackage{colordvi}
\usepackage{color}
\usepackage{dcolumn,amsmath,amsthm,amscd,amsfonts,amssymb,epsfig,graphics,graphicx,eucal}

\newcommand{\be}{\begin{equation}}
\newcommand{\ee}{\end{equation}}
\def\bea{\begin{eqnarray}}
\def\eea{\end{eqnarray}}

\def\phi{\varphi}

\def\bc{\begin{center}}
\def\ec{\end{center}}

\def\Re{\mathrm{Re}}
\def\Im{\mathrm{Im}}

\begin{document}

\title{BIC in waveguide arrays within a symmetry classification scheme}
\author{J. Petr{\' a}{\v c}ek$^{1,2}$,  and  V. Kuzmiak$^{3}$}
\affiliation{
$^1$Institute of Physical Engineering, Faculty of Mechanical Engineering, Brno University of Technology, Technick\'a 2, 616 69 Brno, Czech Republic
\\
$^2$Central European Institute of Technology, Brno University of Technology, Purky{\v n}ova
656/123, 612 00, Brno, Czech Republic\\
$^3$Institute of Photonics and Electronics, Academy of Sciences of the Czech Republic,v.v.i., Chabersk{\'a} 57, 182 51, Praha 8, Czech Republic
}

\date{\today}

\begin{abstract}
We study a photonic analog of a modified Fano-Anderson model -- a waveguide array with two additional waveguides and by using the coupled mode theory we calculate its spectral and scattering properties. We classify eigenomodes according to vertical symmetry of the structure given by self-coupling coefficients of the additional waveguides and establish the conditions for BIC existence. We use the Weierstrass factorization theorem and interpret the scattering spectra of the systems with broken symmetry in terms of the eigenmodes.
The Fano resonance related with excitation of quasi-BIC is explained as arising from the interference between this mode and another leaky mode.
\end{abstract}
\maketitle

\input{epsf.tex}
\epsfverbosetrue

A counterintuitive concept of bound states in the continuum (BICs) was initially proposed
by von Neumann and Wigner \cite{neumann} as a mathematical construct represented by special states which lie in the continuum but remain localized without any radiation as a consequence of
confining potential.
By inspecting various scenarios of light scattering and their relation to location of the poles and zeros of the scattering matrix, BICs were identified as the limiting case of general Hermitian scatterer, when the pole and zero coalesce on the real axis with mutual coherent destruction \cite{monticone}.
BICs have been found in a wide range of different material platforms and have been experimentally observed in electromagnetic, acoustic and water waves
\cite{Hsu}-\cite{ysk}.

\label{sec-1}
\begin{figure}[t]
\centering
\includegraphics[width=8.3cm,clip]{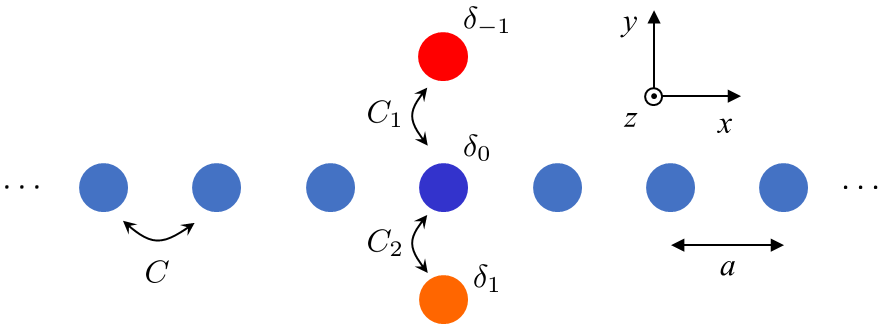}
\caption{A coupled system consisting of a periodic array with two additional waveguides above and below
the array.}
\label{fig-1}       
\end{figure}

In this Letter we employ a theoretical model \cite{PK_pra} to investigate the coupled system consisting of an infinite array of single-mode optical waveguides, with two additional waveguides above and below the array (Fig.~1).
Our configuration can be viewed as an extension of a photonic implementation of the Fano-Anderson model
\cite{Mahan} with a single additional waveguide \cite{ysk2}.
This structure can be regarded as generalization of a famous finite system in which an optical symmetry-protected BIC was confirmed experimentally \cite{plotnik}.
Surprisingly, its comprehensive theoretical understanding has not yet been presented.
We note that a similar structure consisting of a discrete linear chain, which in addition to Fano defects contains one $\delta$-like defect which gives rise to the Fano-Feshbach resonance, was considered in \cite{nguyen}.

To explore physics and to provide a theoretical analysis of the formation of the symmetry-protected BIC, we determine analytically and calculate numerically spectrum of eigenmodes as well as scattering properties of the waveguide array.
We establish the conditions, given by the relation between the self-coupling coefficients associated with side-coupled waveguides $\delta_{\pm 1}$ and the change of the propagation constant $\varepsilon$, that define regimes in which the governing coupled dynamic equations support the different eigenmode spectra.
The eigenvalues $\varepsilon$ reveal new and interesting features, such as the level repulsion and exceptional points
(EPs), that have been found by tracing their position within the complex plane with increasing difference between
$\delta_1$ and $\delta_{-1}$.
All the modes supported by the structure manifest themselves in the scattering spectra as the resonances that can be identified by means of the generalized Weierstrass factorization approach \cite{bonod1,bonod2}. We found that the resonance associated with quasi-BIC state possesses a Fano line-shape arising from the interference between quasi-BIC and another leaky mode.
The transmittance near the resonance can be rewritten into the form of the Fano formula, where the shape parameter $f$ can be expressed in terms of the poles associated with the interacting modes. The knowledge of a set of discrete singular points associated with the scattering matrix, provides a widely applicable framework to interpret and to engineer the resonant properties of the nanostructures \cite{bonod1,bonod2}.

{\em I. Model.}
To describe the coupled structure (Fig. 1) we use the standard approach based on the coupled mode theory (CMT) \cite{hardy,shi}. The total field is represented with slowly varying modal amplitudes
$\psi_m(z)$ ($m\in Z$) and $\phi_{\mp 1}(z)$ on the $m-$th waveguide in the array and on the upper/lower additional waveguide, respectively.
The evolution of the amplitudes is described by a set of coupled equations
\begin{gather}
i\psi_m' = C\left(\psi_{m-1} + \psi_{m+1}\right),\; m\neq 0 \label{eq_CMT_1} \\
i\psi_0' = \delta_{0} \psi_{0}+C\left(\psi_{-1} + \psi_{1}\right) +
C_1\varphi_{-1}+C_2\varphi_{1} \;\label{eq_CMT_2}\\
i\varphi_{-1}' = \delta_{-1} \varphi_{-1} +  C_1\psi_{0} \label{eq_CMT_3}\\
i\varphi_{1}' = \delta_{1} \varphi_{1} +  C_2\psi_{0}, \label{eq_CMT_4}
\end{gather}
where the prime stands for the derivative according to the $z$, $C$ is the coupling coefficient in the array, $C_{1,2}$ are the coupling coefficients between the upper/lower additional waveguide and $0$-th waveguide in the array,
and $\delta_{0}$ and $\delta_{\mp 1}$ is self-coupling coefficients in
$0$-th waveguide in the array and in the upper/lower additional waveguide, respectively.
The self-coupling coefficients may include the effect of nearest-neighbours due to the geometry as well as the possible external perturbations of $0$-th waveguide and the additional waveguides. We assume
that all the parameters which enter Eqs.~(\ref{eq_CMT_1})-(\ref{eq_CMT_4}) are real.

The unperturbed array supports propagation of Bloch waves $\psi_m=A\exp\left(-i k_xam-i\varepsilon z\right)$,
where $k_x$ is the Bloch wavenumber, $a$ is the period of the lattice and the ``energy''
(change of the propagation constant) $\varepsilon$ fulfils the well-known dispersion relation
\be
\varepsilon = 2C\cos\left( k_x a\right).  \label{eq_disp_relation}
\ee

\begin{figure}[th]
\centering
\includegraphics[width=8.6cm,clip]{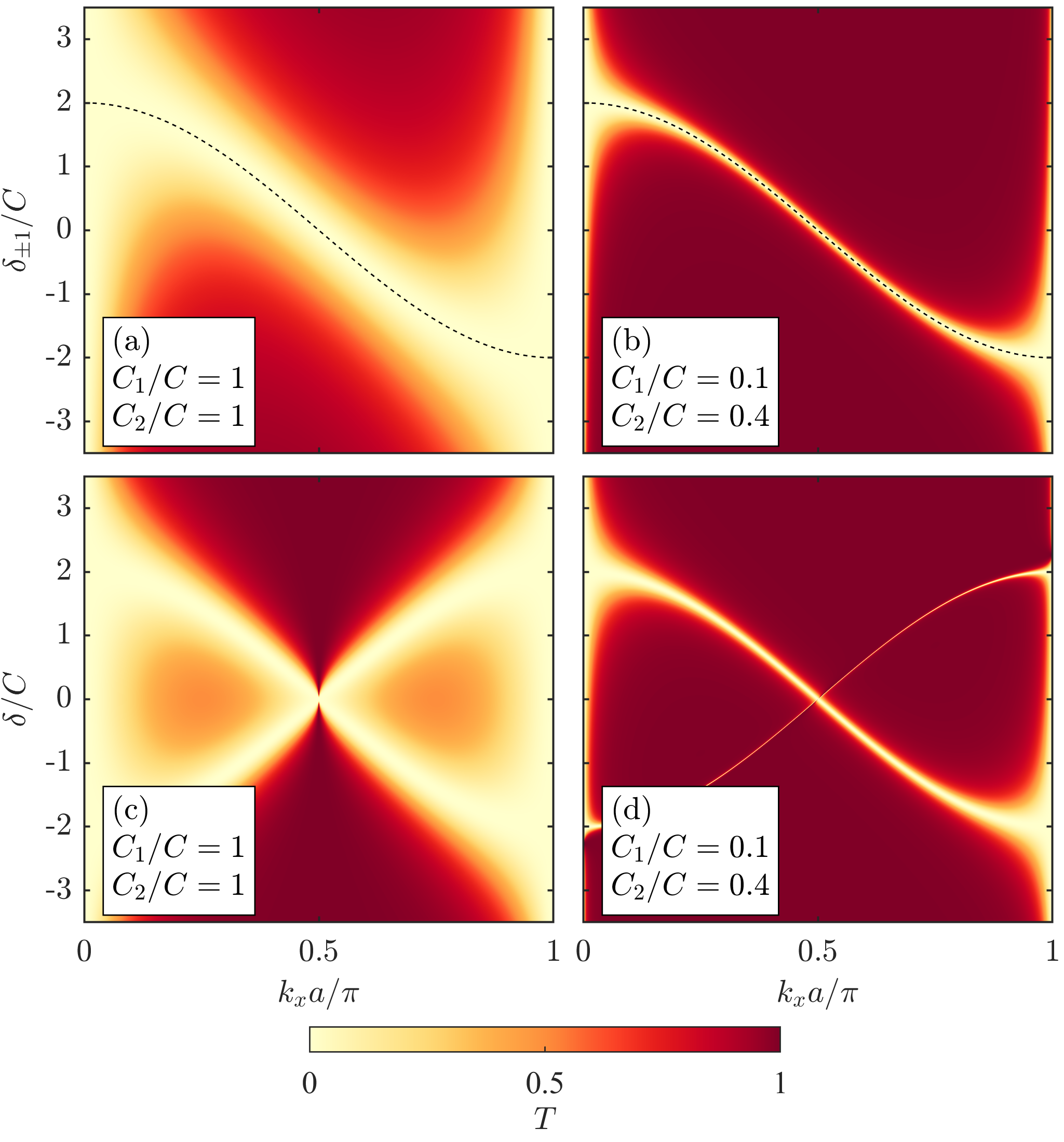}
\caption{The transmittance $T$ of the waveguide array. (a), (b) systems with the symmetry
$\delta_{-1}= \delta_1$;
(c), (d) asymmetric systems 
defined as
$\delta=-\delta_{-1}=\delta_1$; $k_x$ is the Bloch wavenumber, $C_{1,2}$
are shown in the boxes, $\delta_0=0$.
The dashed line in (a), (b) displays the dispersion relation (\ref{eq_disp_relation})
with $\delta_{\pm 1} = \varepsilon $.}
\label{fig-T_map}       
\end{figure}

{\em II. Scattering problem.}
The additional waveguides cause scattering of the Bloch waves.
To evaluate the effect, we follow the standard approach, see, e.g., \cite{Weimann,ysk2}, and represent the field in the left part of the periodic array in terms of incident and reflected waves
\be
\psi_m=e^{-i k_xam-i\varepsilon z} + r e^{i k_xam-i\varepsilon z}, \; m < 0 \label{wave_left}
\ee
in the right part as the transmitted wave
\be
\psi_m=t e^{-i k_xam-i\varepsilon z},  \; m > 0.   \label{wave_right}
\ee
and the fields in the additional waveguides as
\be
\phi_{\pm 1}=B_{\pm 1} e^{-i\varepsilon z}  \label{eq_bs_phi}
\ee
After substituting Eqs.~(\ref{wave_left})-(\ref{eq_bs_phi}) into Eqs.~(\ref{eq_CMT_1})-(\ref{eq_CMT_4}) and few straightforward steps
one finds that Eqs.~(\ref{wave_left}), (\ref{wave_right}) are valid also for $m=0$ and obtains the analytical expressions for the amplitude transmittance $t$ and reflectance $r$ as
\be
t = r+1=
 \frac{2C\sin(k_xa)}{2C\sin(k_xa)+i \mu}  \label{eq_t}
\ee
where $\mu$ reads
\be
\mu = \delta_0 + \frac{C_1^2}{\varepsilon-\delta_{-1}} + \frac{C_2^2}{\varepsilon-\delta_{1}}.   \label{eq_def_mu}
\ee

Figure~\ref{fig-T_map} demonstrates typical behaviour of the intensity transmittance $T = |t|^2$ evaluated for various structural parameters. Figs.~\ref{fig-T_map} (a), (b) show results for {\em symmetric structures} where the self-coupling coefficients in both additional waveguides are the same $\delta_1 = \delta_{-1}$.
The most dominant feature is vanishing $T$ at single point $\varepsilon = \delta_{\pm 1}$
(for fixed $k_x$, $0<k_xa<\pi$) due to diverging parameter
$|\mu|\rightarrow \infty$ in Eq.~(\ref{eq_t}), see also Eq.~(\ref{eq_def_mu}).
This resonance corresponds to existence of the symmetry protected BIC and, for varying $k_x$, is indeed located at positions given by the dispersion relation (\ref{eq_disp_relation}) [see dashed lines in Figs.~\ref{fig-T_map} (a), (b)].
Note also, that the zero transmittance at $\varepsilon = \delta_{\pm 1}$
is double degenerate and identical with energies of localized states of the additional waveguides, i.e., when $\varepsilon$ is calculated from Eqs.~(\ref{eq_CMT_3}) and (\ref{eq_CMT_4}) for $C_{1,2}=0$.

\begin{figure}[th]
\centering
\includegraphics[width=8.6cm,clip]{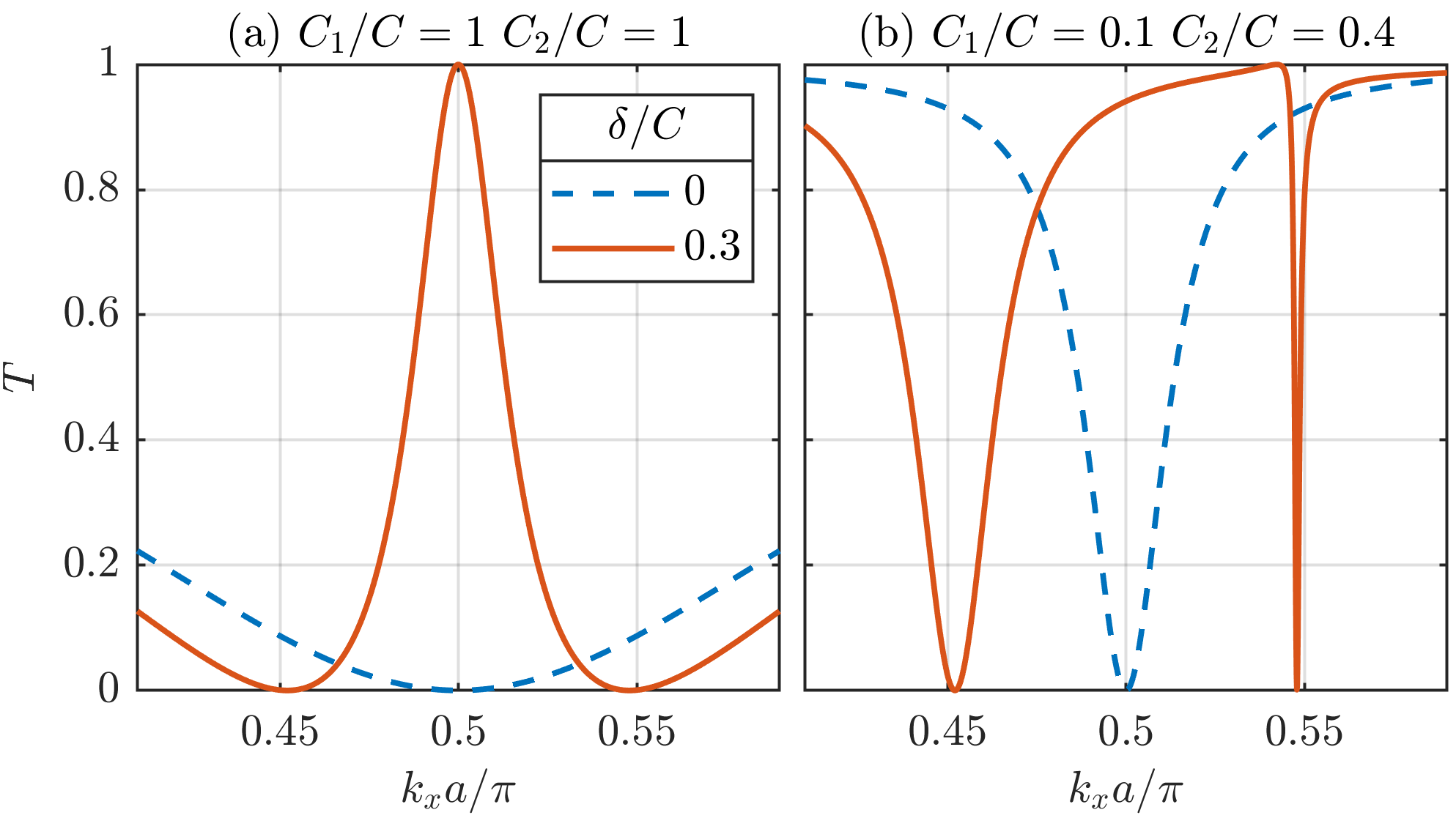}
\caption{The transmittance $T$ vs. the Bloch wavenumber $k_x$ for systems with preserved $\delta=0$ (dashed line) and broken $\delta/C=0.3$
(solid line) symmetry for the coupling coefficients
indicated in the boxes
and $\delta_0=0$.}
\label{fig-Tcurves_break}       
\end{figure}

By breaking the vertical symmetry, the degeneracy is lifted as is illustrated in Figs.~\ref{fig-T_map} (c), (d) and
Fig.~\ref{fig-Tcurves_break} for
{\em asymmetric structures} defined as $\delta=-\delta_{-1}=\delta_1$, where the parameter $\delta$ describes an {\em asymmetry strength}.
We observe two different zeros with energies equal to those of states of isolated additional waveguides,
$\varepsilon = \pm\delta$. It will be shown further, that the resonances correspond to the excitation of two leaky modes of the composed structure, one of them being quasi-BIC, and the related spectral features may be explained by interference of the two leaky modes that leads to the Fano profile.

Importantly, the zero positions observed in Figs.~\ref{fig-T_map}, \ref{fig-Tcurves_break} do not depend on the coupling coefficients
$C_{1,2}$. On the other hand, $C_{1,2}$ have profound influence on the shape of the resonances:
decreasing of $C_1/C$ and $C_2/C$ (weaker coupling) leads to narrower dip in the transmittance $T$,
while asymmetry in $C_{1,2}$ strongly affects asymmetric behaviour of the two dips [compare
Figs.~\ref{fig-Tcurves_break} (a) and (b)].

{\em III. Modes.}
The spectral features described above can be explained in terms of the modes of the composed structure. The modal field in the linear chain has the form
\be
\psi_m=A e^{- i k_xa|m|-i\varepsilon z}, \label{eq_bs_psi}\\
\ee
while that in the additional waveguides is given by Eq.~(\ref{eq_bs_phi}).
By substituting Eqs.~(\ref{eq_bs_psi}) and (\ref{eq_bs_phi}) into Eqs.~(\ref{eq_CMT_2})-(\ref{eq_CMT_4}) we obtain the following nonlinear eigenvalue problem
\begin{gather}
\left[\delta_0-\varepsilon+2C\exp\left(- ik_x a\right)\right]A+C_1B_{-1}+C_2B_1=0, \label{eig_A}\\
C_1A+ \left(\delta_{-1}-\varepsilon\right)B_{-1}=0, \label{eig_Bm1}\\
C_2A+ \left(\delta_{1}-\varepsilon\right)B_{1}=0.\label{eig_Bp1}
\end{gather}
The relation between the self-coupling coefficients $\delta_{\pm 1}$ and change of the propagation constant
$\varepsilon$ plays the key role
in a symmetry classification scheme, which defines the regimes with different families of the eigenmodes.

{\em A.}
We first consider the case when $\varepsilon=\delta_1$ or $\varepsilon=\delta_{-1}$.
According to Eqs.~(\ref{eig_A})-(\ref{eig_Bp1}) a nontrivvial solution is not supported unless {\em the structure possesses the vertical symmetry}
$\delta_{-1}=\delta_1$ when one obtains a doubly-degenerate bound state $\varepsilon = \delta_{\pm 1}$, which yields
{\em a vertically antisymmetric eigenfunction}
$C_1B_{-1}=-{C_2}B_1$, $A = 0$ and represents the symmetry protected BIC provided $-2C<\delta_{\pm 1}<2C$.

{\em B.}
Now we examine the case $\varepsilon\neq\delta_{1}$ and $\varepsilon\neq\delta_{-1}$.
Eqs.~(\ref{eig_Bm1})-(\ref{eig_Bp1}) yield nontrivial solutions for the amplitudes
associated with the attached waveguides in the form
\be
B_{\mp 1}=\frac{C_{1,2}}{\varepsilon - \delta_{\mp 1}}A \label{eq_B_ampl}.
\ee
By substituting the both expressions into Eq.~(\ref{eig_A}) one obtains the eigenvalue equation
\be
2C\sin(k_xa)+ i\mu = 0.
\label{eq_eig}
\ee
It is important to keep in mind, that none of the solutions of Eq.~(\ref{eq_eig}) is a true BIC as we have $A\neq 0$, however, for the vertically asymmetric structures ($\delta_{-1}\neq \delta_1$), one of them may represent a quasi-BIC as will be shown below.

{\em B.1 Symmetric states.}
The family of the symmetric states can be divided to several groups according to the vertical symmetry of the structure.
{\em B.1a} We first consider the structure with {\em the trivial vertical symmetry},
$\delta_{-1}= \delta_1 = \delta_0=0$, for which Eq.~(\ref{eq_eig}) provides 4 eigenvalues
\be \label{eq_epsilon_trv}
\varepsilon = \pm \sqrt{2C^2 \pm 2\sqrt{C^4+C_{\rm av}^4}},
\ee
where all combinations of the signs are allowed and $C_{\rm av}^2 = (C_1^2+C_2^2)/2$.
The modes (in contrast to antisymmetric modes supported by the same vertically symmetric structures) exhibit symmetry
\be
C_2B_{-1}={C_1}B_1.  \label{eq_vert_sym}
\ee

\begin{figure}[th]
\centering
\includegraphics[width=8.6cm,clip]{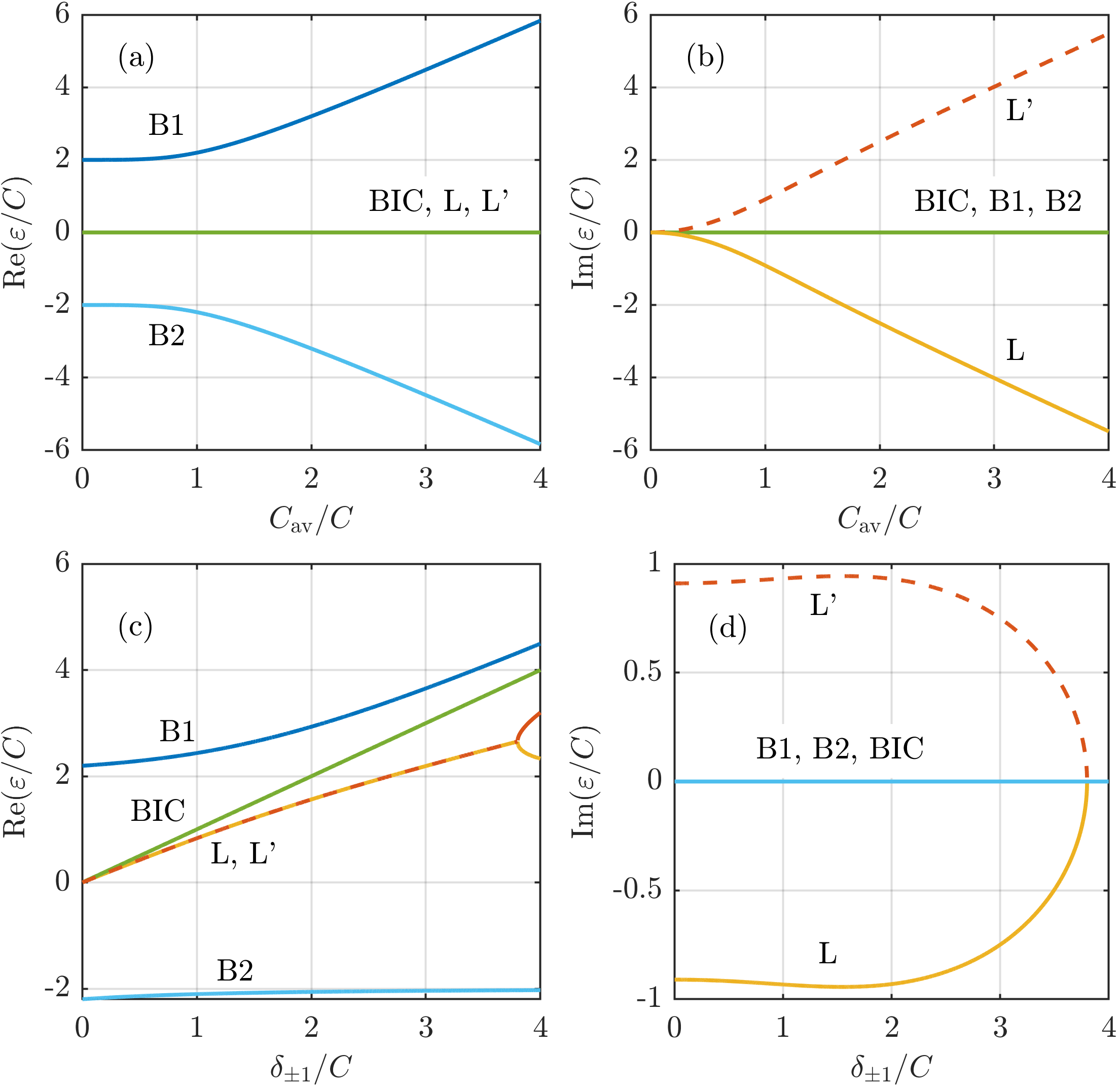}
\caption{(a),(b) Spectrum of structures with trivial, $\delta_{\pm 1}=\delta_0=0$, and
(c),(d) nontrivial, $\delta_{-1}=\delta_{1}$, vertical symmetry. (a) The real and (b) imaginary part of the energy
$\varepsilon$ as a function of $C_{\rm av}/C$.
(c) The real and (d) imaginary part of $\varepsilon$ as a function of
$\delta_{\pm 1}/C$ for $\delta_0=0$ and $C_1=C_2=C$.
The modes in (c), (d) are labelled according to their asymptotics  when $\delta_{\pm 1} \rightarrow 0$.
Dashed lines indicate the states with $\Im(\varepsilon)>0$.
}
\label{fig-eps_trv}       
\end{figure}

\begin{figure}[th]
\centering
\includegraphics[width=8.6cm,clip]{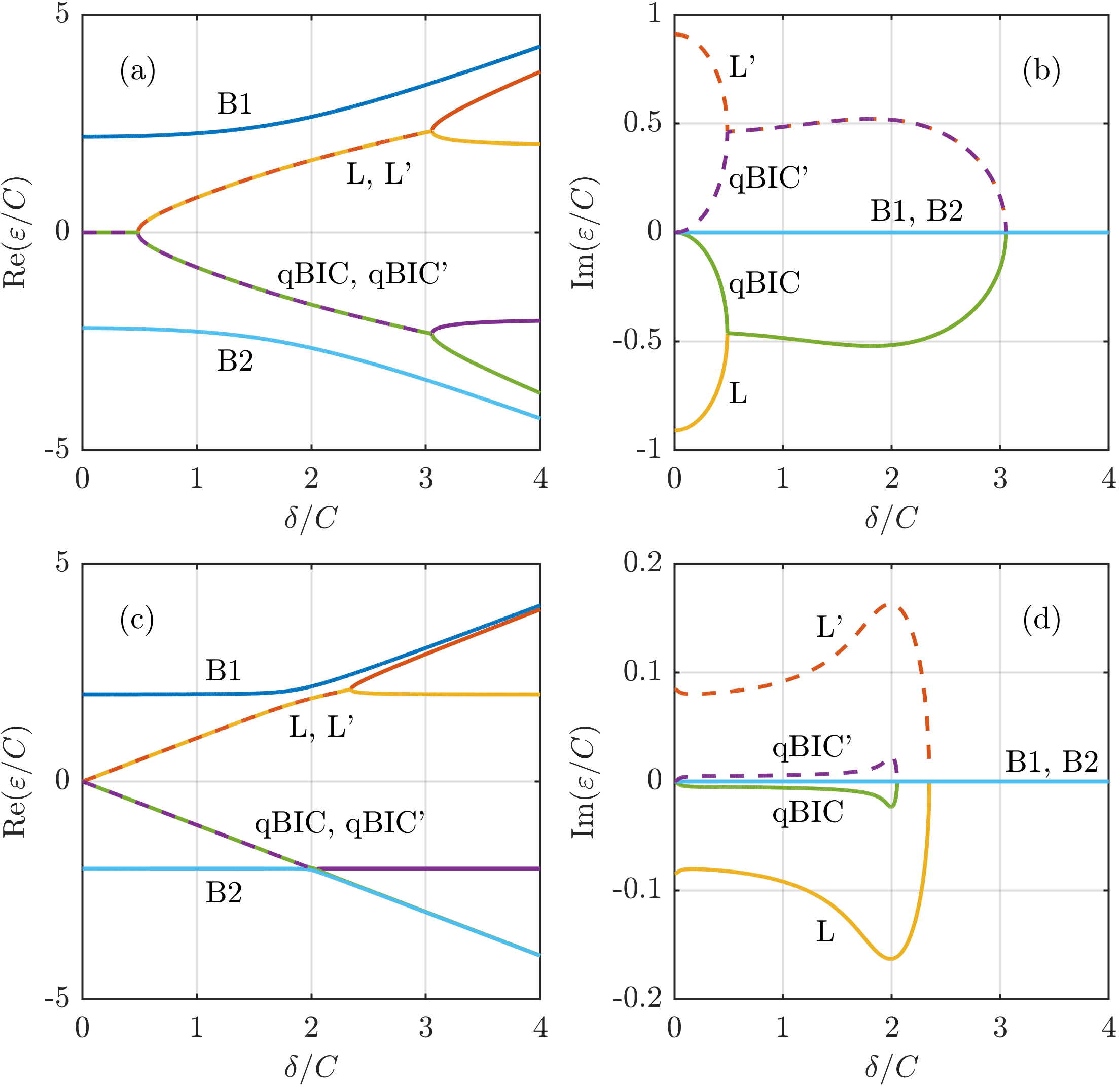}
\caption{Spectrum of structures with broken vertical symmetry $\delta=-\delta_{-1}=\delta_1$:
the energy $\varepsilon$ vs. the asymmetry strength $\delta$. (a), (b)  $C_1=C_2=C$;
(c), (d) $C_1/C=0.1$, $C_2/C=0.4$; $\delta_0=0$.
}
\label{fig-eps_delta}
\end{figure}

The full spectrum of the trivially symmetric structures, shown in Fig.~\ref{fig-eps_trv}(a),(b), consists of symmetry protected doubly-degenerate BIC within the continuum, two bound modes (B1,B2) with real energy, which bifurcate from the lower and upper edges of the continuum when the parameter $C_{\rm av}$ is increased, and two modes (L,L') with pure imaginary energy.
We note that Eq.~(\ref{eq_epsilon_trv}) illustrates the general feature of Eq.~(\ref{eq_eig}): the energy eigenvalues are either real or occur in complex conjugate pairs. The states with $\Im(\varepsilon)<0$ (solid lines in Figs.~\ref{fig-eps_trv} and
\ref{fig-eps_delta}) fulfil the conventional definition of leaky modes and satisfy outgoing boundary conditions
$0<\Re(k_xa)\le\pi$, states with $\Im(\varepsilon)>0$ (dashed lines in Figs.~\ref{fig-eps_trv} and \ref{fig-eps_delta}) correspond to ingoing boundary conditions $-\pi<\Re(k_xa)<0$ \cite{PK_pra}.

{\em B.1b}
For the structure belonging to the class characterized by {\em the general vertical symmetry} $\delta_{-1}= \delta_1 \neq 0$, Eq.~(\ref{eq_eig}) yields 4 solutions, the modes are symmetric according to
Eq.~(\ref{eq_vert_sym}).
The spectrum shown in Fig.~\ref{fig-eps_trv}(c),(d) also consists of the bound modes
(B1, B2, BIC); the latter one enters the gap at $\delta_{\pm 1}/C =2$ when $\delta_{\pm 1}/C$ is increased.
The L, L' modes initially exhibit complex conjugate energies $\varepsilon$, the real component of their energy $\Re(\varepsilon)$ also enters the gap at $\delta_{\pm 1}/C = 2.5$. The imaginary component of their energy $\Im(\varepsilon)$ remains nonzero in the range $2.5 < \delta/C < 3.1$ until the (L,L') doublet reaches the EP beyond which the level repulsion effect occurs. Here the level repulsion, in contrast to the $\cal{PT}$-symmetric structures, arises from the coupling between two degenerate modes with positive energy \cite{bernier}.

{\em B.2 Structures with broken vertical symmetry (asymmetric states).}
Breaking the vertical symmetry, $\delta_{-1}\neq \delta_1$, yields coupling of the true BIC antisymmetric state to the continuum and the doubly degenerate mode is split into quasi-BIC (qBIC) and qBIC' mode as shown in
Fig.~\ref{fig-eps_delta}, which demonstrates a rich spectral behavior of the structure. Eq.~(\ref{eq_eig}) yields 6 solutions; instead of Eq.~(\ref{eq_vert_sym}), the amplitudes $B_{\pm 1}$ satisfy the condition $C_2B_{-1}(\delta_{-1}-\varepsilon)= C_1B_1(\delta_1-\varepsilon)$.

The spectra for the structures with $C_1=C_2$, see Figs.~\ref{fig-eps_delta}(a),(b), reveal in comparison with those shown in Fig.~\ref{fig-eps_trv} the EP which occurs when asymmetry strength $\delta/C = 0.5$.
The behavior of the EP differs from common one known in $\cal{PT}$-symmetric structures:
the L, L' and qBIC, qBIC' doublets, initially (for $\delta/C < 0.5$)
with complex conjugate values of energy $\varepsilon$ and vanishing its real component,
$\Re(\varepsilon)=0$, coalesce at the EP in L, qBIC and L', qBIC' pairs.
In the range between $0.5 < \delta/C < 3.1$, the common value of $\Re(\varepsilon)$
for the L, L' (or qBIC, qBIC') doublet increases (decreases) and approaches the upper (lower) edge of the continuum, while 
$\Im(\varepsilon)$ for the L, qBIC and L', qBIC' pairs coalesce.
Beyond another EP at $\delta/C = 3.1$ the L, L' and qBIC, qBIC' doublets split into the singlets and transform into the bound states.

The structures with nonequal coupling coefficients $C_1\neq C_2$ exhibit, when increasing $\delta/C$ from zero
thresholdless behavior, see Fig.~\ref{fig-eps_delta}(c),(d), i.e., the common value of $\Re(\varepsilon)$
for the L, L' (or qBIC, qBIC') doublets starts to increase
(decrease) with $\delta/C$ and enters the gap above (below) the continuum.
With further increasing $\delta/C$
the doublets L, L' and qBIC, qBIC' reach EPs at which they also transform into the bound states with $\Im(\varepsilon)=0$.

\begin{figure}[th]
\centering
\includegraphics[width=8.6cm,clip]{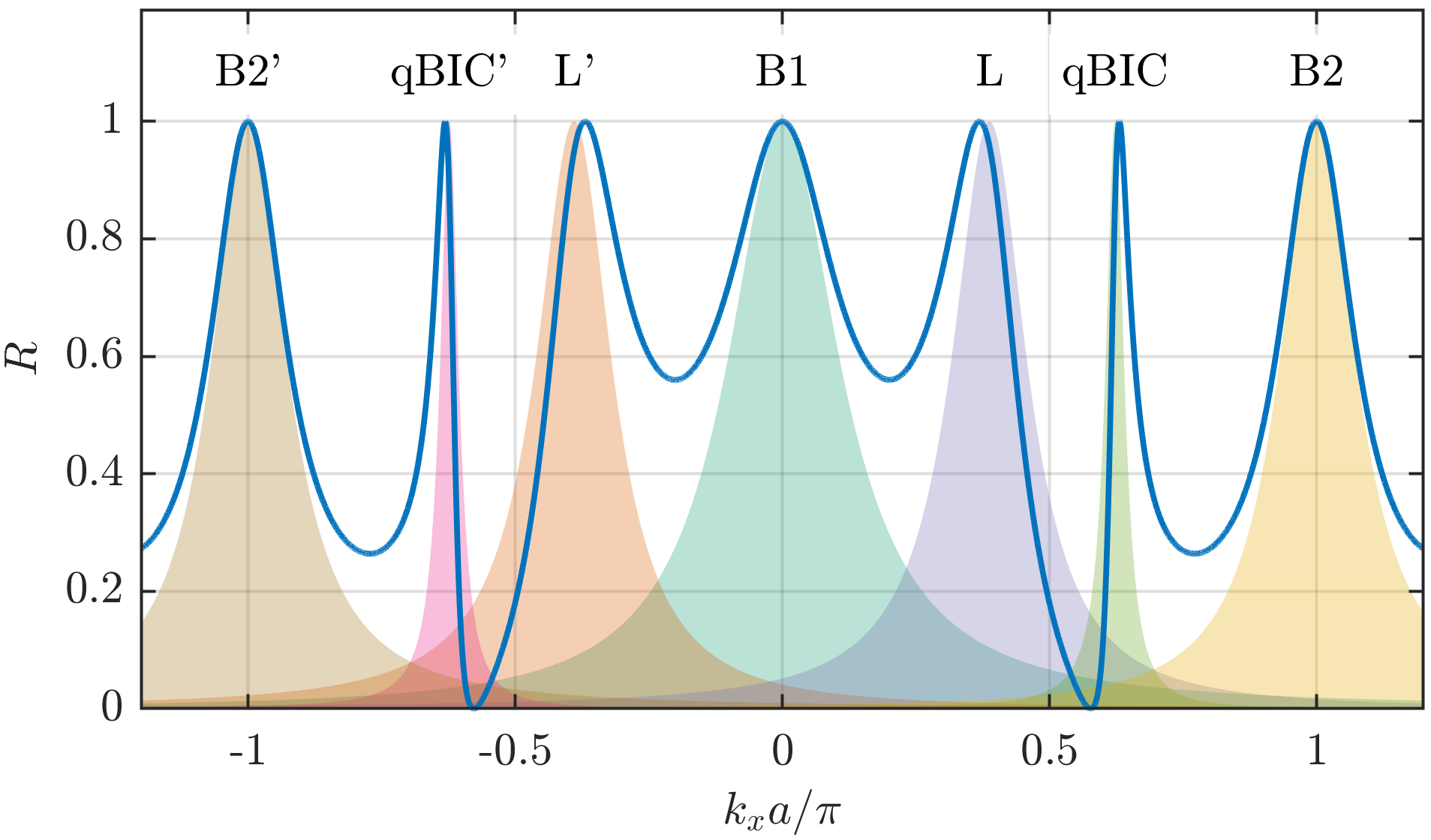}
\caption{The reflectance $R$ vs. the Bloch wavenumber $k_x$ for the system with 
$C_1 = 0.5C$, $C_2 = C$,
$\delta=0.8C$
and $\delta_{0}=0$.
The shaded areas correspond to the calculations when only the indicated pole is used in the
expansion given by Eq.~(\ref{eqFactorization}).}
\label{fig-WF1}       
\end{figure}

{\em IV. Factorization approach.}
The scattering properties of an arbitrary structure can be characterized by the scattering matrix,
which for our structure
has two eigenvalues \cite{PK_pra} $S_o=r-t=-1$ and
\be
S_e=r+t=2t-1=\frac{2C\sin(k_xa)-i \mu}{2C\sin(k_xa)+i \mu}.  \label{eq_S_e}
\ee
The dependence $S_e(q)$, $q\equiv k_xa$, can be expressed through the Weierstrass factorization theorem \cite{bonod1,bonod2} which, for our structure, takes the simple
form \cite{PK_pra}
\be
\label{eqFactorization}
S_e(q) =  \prod\limits_n {\frac{q - p_n^\ast}{q - p_n}},
\ee
where $p_n$ are poles of $S_e(q)$. It follows from
Eq.~(\ref{eq_S_e}) that the poles $p_n$ are solutions of the eigenvalue problem
(\ref{eq_eig}) in terms of $k_xa$. Therefore, the knowledge of eigenmodes is sufficient to restore $S_e(q)$
through Eq.~(\ref{eqFactorization}) and to evaluate the scattering spectra as $r=(S_e-1)/2$
and $t=(S_e+1)/2$, where, in the reflectance $R=|r|^2$, each pole creates a peak of Lorentzian shape \cite{bonod1}.
This way, we can interpret observed spectral features in terms of the eigenmodes. The procedure is illustrated in
Fig.~\ref{fig-WF1} where, in addition to the reflectance (solid line), we also display
contributions of the individual modes (shaded areas) to the total spectrum.
To demonstrate the effect of poles corresponding with all 6 previously described eigenmodes we extended the range of
$k_x$ (B2' refers to the pole B2 shifted by $-2\pi$).
Clearly, B1 and B2 are responsible for peaks at band edges while L and qBIC for the peaks in the continuum.
Even L' and qBIC' (or other poles such as B2') may affect the spectrum
in the physical range $0\le k_xa \le \pi$ provided that the overlap with their tails is sufficiently large.

The destructive interference mainly between L and qBIC modes leads to the asymmetric 
resonance associated with qBIC mode and minima $R=0$. This feature corresponds to the asymmetric dip and maxima $T=1$ in
the transmittance, such as in Fig.~\ref{fig-Tcurves_break} (b).
In fact, it can be shown \cite{PK_pra} that superposition of L and qBIC modes, playing the roles of the continuum and a narrow discrete state
with poles $p_c$ and $p_s$, respectively, leads to the Fano formula
\be \label{eqFanoT}
T(\Omega)=\left(\frac{1}{1+f^2}\right)\frac {\left(\Omega +f\right)^2}{\Omega^2+1},
\ee
where
\be
\Omega \equiv \frac{q-\Re(p_s)}{\Im(p_s)}, \;\;
f \approx \frac{\Im(p_c)}{\Re(p_c)-\Re(p_s)}.
\ee


In conclusion, we studied systematically spectral and scattering properties of a photonic analog of an extended Fano-Anderson tight-binding model.
We classified the modes according to the relation between the self-coupling coefficients and the change of the propagation constant.
The resonant features in the scattering spectra of the structures with broken symmetry can be interpreted through a generalized Weierstrass factorization. In particular, the Fano resonance associated with quasi-BIC arises from the interference between this state and another leaky mode. The transmittance near the quasi-BIC resonance can be rewritten into the form of the Fano formula, where the shape parameter $f$ can be expressed in terms of the poles of the interacting modes. Our work provides the theoretical framework which describes transformation of the symmetry protected BIC into a leaky mode and allows to interpret the resonant properties of the more complex systems. In addition through the engineering of zeros of the transmission matrix, it offers a possibility to investigate their nontrivial topological properties \cite{genack} and may prove to be useful in various fields of optics as well as in cold matter and quantum confined systems.

We acknowledge financial support by the Czech Science Foundation (CSF) through the project 1900062S.

\end{document}